\documentclass[pra,aps,10pt,twocolumn,superscriptaddress,showpacs,floatfix]{revtex4-1}

\usepackage{natbib}

\usepackage{graphicx}
\usepackage{dcolumn}
\usepackage{amsmath}
\usepackage{amssymb}
\usepackage{bm}
\usepackage{afterpage}

\def\beq{\begin{equation}}
\def\eeq{\end{equation}}

\def\om{\omega}
\def\a{\alpha}

\def\s{\sigma}
\def\D{\Delta}

\def\ad{a^\dagger}
\def\rd{{\rm{d}}}

\def\p{\phi}

\def\ra{\rightarrow}

\def\Cc{\mathbb{C}}
\def\Zz{\mathbb{Z}}
\def\Rr{\mathbb{R}}

\def\T{\hat{T}}

\def\Hp{{\cal H}_+}
\def\Hm{{\cal H}_-}

\def\bz{\bar{z}}
\def\tx{\tilde{x}}
\def\pa{\partial}
\def\A{\mathbf{A}}

\def\ch{{\rm ch}}
\def\sh{{\rm sh}}
\def\tp{\tilde{\phi}}
\def\v{\bm{v}}
\def\A{\bm{A}}
\def\B{\bm{B}}
\def\C{\bm{C}}

\newcommand{\expec}[1]{\langle #1 \rangle}

\newcommand{\abs}[1]{\vert #1 \vert}
\newcommand{\ket}[1]{\vert #1 \rangle} 
\newcommand{\bra}[1]{\langle #1 \vert} 
\newcommand{\braket}[2]{\langle #1 \vert #2 \rangle}

\def\hsizeNew{1.0\hsize}

\makeatletter
\def\@listcomma@comma{\@ifnum{\@tempcnta>\@ne}{}{}}%
\def\@listcomma@comma@UK{\@ifnum{\@tempcnta>\tw@}{}{}}%
\makeatother

\begin{document}

  \title{Exact real-time dynamics of the quantum Rabi model}

 \author{F. Alexander Wolf,}
  \affiliation{Experimental Physics VI, $^2$Theoretical Physics III,\\
    Center for Electronic Correlations and Magnetism,\\
    University of Augsburg, 86135 Augsburg, Germany}
  \author{Marcus Kollar,$^2$}
  \noaffiliation
  \author{Daniel Braak}
  \affiliation{Experimental Physics VI, $^2$Theoretical Physics III,\\
   Center for Electronic Correlations and Magnetism,\\
    University of Augsburg, 86135 Augsburg, Germany}

  \date{April 1, 2012}

  \begin{abstract}
    We use the analytical solution of the quantum Rabi model to obtain
    absolutely convergent series expressions of the exact eigenstates
    and their scalar products with Fock states.
    This enables us to calculate the numerically exact time evolution
    of $\expec{\sigma_x(t)}$ and $\expec{\sigma_z(t)}$ for all regimes
    of the coupling strength, without truncation of the Hilbert space.
    We find a qualitatively different behavior of both observables
    which can be related to their representations in the invariant
    parity subspaces.  
  \end{abstract}


  \maketitle

  \section{Introduction}

  The quantum Rabi model (QRM) describes one of the simplest strongly
  coupled quantum systems: a single bosonic mode interacts with a
  two-level system.  It forms the basic building block of theoretical
  approaches to the interaction of light with matter~\cite{allen} or
  to electron-phonon interaction~\cite{rongsheng}. Due to recent
  progress in nanofabrication, the strong-coupling regime of the QRM
  could be accessed experimentally within circuit
  QED~\cite{niemc,forn}, where the two-level system corresponds to a
  single qubit.  To construct controllable solid-state quantum gates,
  it is necessary to comprehend the behavior of the QRM for all regimes
  of the coupling strength~\cite{pell}.

  Although the lower part of the spectrum of the QRM can be computed
  numerically to arbitrary precision using exact diagonalization on a
  truncated state space due to the convergence properties of the
  associated continued fraction~\cite{schweber,nusser,reik,klenner},
  it is by no means evident whether the properties of the numerically obtained
  eigenstates correspond to those of the true eigenstates of the
  untruncated model. Even if the spectra of the truncated and the
  full model are close for a subset of the eigenvalues, the corresponding 
  eigenvectors  
  could still be related by a unitary transformation, which does not
  necessarily approach the identity.  
  Consequently, a variety of systematic approximations
  making use of different basis sets in the infinite-dimensional Hilbert
  space have been developed~\cite{rongsheng,per,pan,liu,grifoni,feranchuk,albert}.
  However, all of them implement a truncation procedure at some stage
  of the calculation and thus cannot answer the basic question on the
  relation of the truncated to the full model. 
   
  Beside this theoretical goal, there is a quite practical reason to investigate
  the true eigenbasis of the QRM:   
  To compute the dynamics of the system on all time scales, not only
  the exact eigenvalues, but also the exact eigenstates  are required.
  Only then one may hope to predict reliably the behavior of
  qubits within a circuit QED setup~\cite{pell}.
    
  In this paper we solve the problem by
  means of the recently obtained exact solution of the QRM~\cite{db}.
  We construct absolutely convergent series expansions of the
  eigenstates in terms of known basis sets, which allows to estimate
  the true error in any numerical evaluation. It turns out that
  a standard calculation with double precision, sufficient to
  compute the spectrum, fails for the eigenstates.  

  The QRM Hamiltonian reads
  \begin{align}
    H_{\rm{R}}=\om\ad a +g\s_x(a+\ad) + \frac{\om_0}{2}\s_z .
    \label{hamr}
  \end{align}
  The Hilbert space is given by
  $L^2(\Rr)\otimes\Cc^2$, where the basis of the spin part of the
  wavefunction $\ket{\s}$ is $\{\ket{+1},\ket{-1}\}$.  We employ the
  Bargmann space $\cal B$ of analytical functions to represent
  elements of $L^2(\Rr)$~\cite{ba}. In $\cal B$ the creation
  (annihilation) operators $\ad$ ($a$) are represented as $z$
  ($\pa_z$) and normalization is defined with respect to the scalar
  product
  \begin{align}
    \langle\psi|\phi\rangle =\frac{1}{\pi}\int\rd z\rd \bz e^{-z\bz}\overline{\psi(z)}\phi(z).
    \label{scalarProd}
  \end{align}
  An arbitrary analytical function $\p(z)$ 
  is an
  element of the Bargmann space $\cal B$ if $\langle\p|\p\rangle <
  \infty$.

  Each eigenstate of $H_{\rm{R}}$ belongs to one invariant subspace (parity
  chain)~\cite{sol} with a fixed $\Zz_2$ quantum number, the parity,
  taking values $\pm 1$.
  The parity operator for $H_{\rm{R}}$ is defined as
  $\hat{P}=(-1)^{z\pa_z}\s_z$.  Due to $[H_{\rm R},\hat{P}]=0$, the
  Hilbert space is a direct sum of invariant subspaces $\Hp\oplus\Hm$
  and $H_{\rm{R}}$ takes the form $H_\pm$ on ${\cal H}_\pm$.

  The invariant parity chains ${\cal H}_\pm$ are spanned by
  $\{\p_s(z)\otimes|\pm 1\rangle,$ $\p_a(z)\otimes|\mp 1\rangle\}$,
  where $\p_{a,s}=(1/2)[\p(z)\mp\p(-z)]$ is the (anti-)symmetric part
  of $\p(z)$.  Each parity chain is therefore isomorphic to $\cal B$
  via the mapping
  \begin{align}
    \p(z) \leftrightarrow 
    \p_s(z)\otimes|\pm 1\rangle+\p_a(z)\otimes|\mp1\rangle,
    \label{isoP}
  \end{align}
  where $\p_s(z)\otimes|\pm 1\rangle$ $+$ $\p_a(z)\otimes|\mp1\rangle$
  $=$ $\ket{\p,\pm}$ is an element of ${\cal H}_\pm$.  Using this
  isomorphism, the Hamiltonian $H_\pm$ reads
  \begin{align}
    H_\pm = \om z\partial_z +g(z+\partial_z)\pm \D\T
    \label{hampar}
  \end{align}
  with $\D=\om_0/2$, and the reflection operator $\T$ acts on elements of
  $\cal B$ as $\T(\p)(z)=\p(-z)$. For notational simplicity, we
  restrict ourselves to $\Hp$ and measure all
  energies in units of $\om$.

  The eigenfunctions $\psi_m(z)$ and eigenvalues $E_m$ of $H_+$ have
  been obtained in Ref.~\onlinecite{db} as two equivalent representations,
  \begin{subequations}%
    \begin{align}%
      \psi_m(z) & = \p_2(x_m,-z) =  e^{gz}\sum_{n=0}^\infty K_n(x_m)(-z+g)^n,
      \label{phi2}\\
      \psi_m(z) & = \p_1(x_m,z)  =  e^{-gz}\sum_{n=0}^\infty K_n(x_m)\D\frac{(z+g)^n}{x_m-n},
      \label{phi1}%
    \end{align}%
    \label{phi}%
  \end{subequations}%
  where $x_m=E_m+g^2$ and
  \begin{subequations}\label{eqkn}%
    \begin{align}%
      n K_n(x) & = f_{n-1}(x) K_{n-1}(x)-K_{n-2}(x),\\
      K_1(x)   & = f_0(x), ~~~    K_0(x)= 1,\\
      f_{n}(x) & = \frac{2g}{\om} + \frac{1}{2g}\Big(n \om -x + \frac{\Delta^2}{x-n \om} \Big).
    \end{align}%
  \end{subequations}%
  Both
  \eqref{phi2} and \eqref{phi1} converge for all $z$ in the complex
  plane if and only if $x_m$ satisfies the spectral condition \cite{db}
  \begin{align}
    & G_+(x_m,0) = 0 \quad\rm{where} \\
    & G_+(x,z) =\p_2(x,-z)-\p_1(x,z).
    \label{gfunction}
  \end{align} 
  For all other values of the spectral parameter $x\neq x_m$, the functions
  $\p_{1,2}(x,z)$ are defined via (\ref{phi2},\ref{phi1}) only within
  the circle $|z+g|<2g$.  Furthermore the eigenfunctions are not
  normalized when expressed in terms of (\ref{phi2},\ref{phi1}). Both
  difficulties are overcome in the following.

  \section{Time evolution}
  \label{sec_timeEvol}

  If an initial state $|\p_0,+\rangle$ with fixed parity $p=1$ is
  prepared at time $t=0$, the time evolution takes place within $\Hp$
  and reads
  \begin{align}
    \ket{\p(t)} = e^{-iH_+t} \ket{\p_0} =
    \sum_m e^{-iE_m^+t}  \ket{\psi_m} \frac{\braket{\psi_m}{\p_0}}{\braket{\psi_m}{\psi_m}}
    \label{timeEvol}
  \end{align}
  The task consists in computing $\langle \psi_m|\p_0\rangle$ and with
  that the norms $\langle\psi_m|\psi_m\rangle$ for $m=0,1,2,\ldots$.

  \subsection{Formal solution} 

  Starting from (\ref{phi2},\ref{phi1}) we obtain an expansion of
  $\ket{\psi_m}$ into normalized eigenstates of the shifted harmonic
  oscillator,
  \begin{align}
    \ket{n;g}= \frac{e^{-g^2/2}}{\sqrt{n!}} (z+g)^n e^{-gz},
    \label{shift-osc}
  \end{align}
  as follows:
  \begin{subequations}%
    \begin{align}%
       \ket{\psi_m} & = e^{g^2/2}\sum_{n=0}^\infty \sqrt{n!}\,K_n(x_m)\,(-1)^n          |n;-g\rangle,\\
       \ket{\psi_m} & = e^{g^2/2}\sum_{n=0}^\infty \sqrt{n!}\,K_n(x_m)\,\frac{\D}{x_m-n} |n; g\rangle.
    \end{align}%
    \label{eigenstates01}%
  \end{subequations}%
  From these equations and the fact that the eigenbasis of the shifted
  harmonic oscillator $\{\ket{n;g}\}$ is orthonormal we obtain two
  equivalent representations of the squared norm of the eigenstates:
  \begin{subequations}%
    \begin{align}%
      \langle\psi_m|\psi_m\rangle 
      & =  e^{g^2}\sum_{n=0}^\infty n!\,K_n^2(x_m) 
      \label{norm1a}\\
      & =  e^{g^2}\sum_{n=0}^\infty n!\,\frac{\D^2}{(x_m-n)^2} K_n^2(x_m).
      \label{norm1b}%
    \end{align}%
    \label{norm1}%
  \end{subequations}%
  Using the expansions of the initial state $\p_0$ into shifted
  oscillator states $\p_0=\sum_{n=0}^\infty\a_n|n;-g\rangle =\sum_{n=0}^\infty\beta_n|n;g\rangle$ we have
  \begin{subequations}%
    \begin{align}%
      \langle\psi_m|\p_0\rangle 
      & = e^{g^2/2}\sum_{n=0}^\infty \sqrt{n!}\,K_n(x_m)\,(-1)^n\,{\a}_n,
      \label{scalar2}\\
      & = e^{g^2/2}\sum_{n=0}^\infty \sqrt{n!}\,K_n(x_m)\,\frac{\D}{x_m-n}\,\beta_n.
      \label{scalar1}%
    \end{align}%
    \label{scalar}%
  \end{subequations}%
  The preceding equations provide a formal solution of the time evolution problem of
  \eqref{timeEvol}.

  \subsection{Asymptotic numerics}

\begin{figure}[t!]
    \centering
    \includegraphics[width=\hsize]{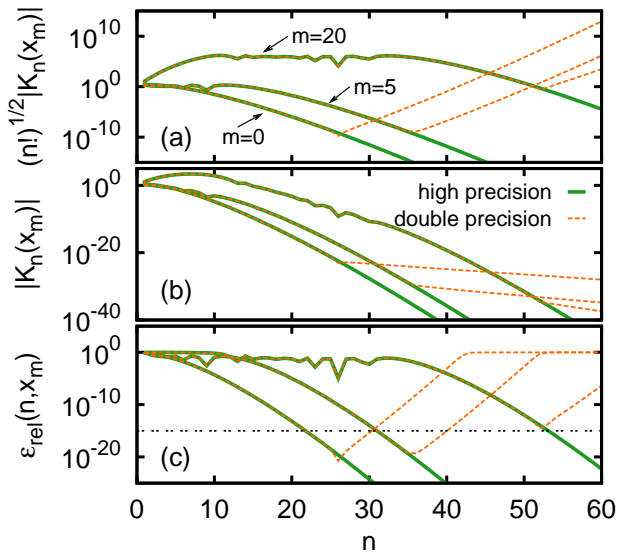}
    \vspace*{-0.3cm}
    \caption{Panel (a) shows the square root of
      the absolute values of the summands $\sqrt{n!}\abs{K_n(x_m)}$ of
      (\ref{norm1a},\ref{norm1approx}).  Panel (b) shows the absolute
      value of the coefficients $\abs{K_n(x_m)}$ (see Eq.~\eqref{eqkn}) of the
      eigenfunctions and panel (c) shows the relative error
      $\epsilon_{\text{rel}}(N=n)$ \eqref{relError} of the series
      expression in Eq.~(\ref{norm1approx}).  All quantities are depicted for
      three eigenenergies $\{x_0, x_5, x_{20}\}$ and the parameters
      $g/\omega=0.7$, $\D/\omega=0.25$.  We present a calculation with double
      precision and $\delta=10^{-10}$ (dashed lines) and a
      calculation with high ($50$ digit) precision and
      $\delta=10^{-30}$ (solid lines). The horizontal
      line in panel (c) marks the value
      $\epsilon_{\text{rel}}=10^{-15}$.}
    \label{fig_Conv}
  \end{figure}

  In practice a numerical calculation of (\ref{norm1},\ref{scalar})
  fails as the sums diverge for all $x$ which do not \textit{exactly}
  coincide with a value $x_m$ of the spectrum.  For every
  approximation $\tilde{x}_m$ to $x_m$, the absolute values of the
  summands of the series in (\ref{norm1},\ref{scalar}) start growing
  for an $n>n'$,  with $n'$ unknown \textit{a priori}.  In
  Fig.~\ref{fig_Conv}(a), the dashed lines show this behavior for the
  example of the summands of the norm calculated via \eqref{norm1a}.
  The picture suggests the numerical finding we confirmed for
  different parameters, that $n'$ grows for growing eigenenergies
  $x_m$.  We further find numerically that $n'$ is large if the error
  of the approximation $\delta = \abs{x_m-\tilde{x}_m}$ is small, which is
  characteristic for series that are asymptotically valid.  It will be
  shown in the next section that this asymptotic validity holds
  rigorously true for the series in (\ref{norm1},\ref{scalar}).

  Due to its asymptotic validity for $\delta$ $\to$ $0$, we can
  evaluate \eqref{norm1a} by approximating
  \begin{align}
    \braket{\psi_m}{\psi_m}
    =
    e^{g^2}\sum_{n=0}^{N} n!K_n^2(\tilde{x}_m) + R_N.
    \label{norm1approx}
  \end{align}
  Here the cutoff $N$ must be chosen large enough so that the relative
  error $R_N/\braket{\psi_m}{\psi_m}$, estimated as
  \begin{align}
    \epsilon_{\rm{rel}}(N) = 
    \frac{N! 
      K_{N}^2(\tilde{x}_m) }{\sum_{n=0}^{N-1} n! K_n^2(\tilde{x}_m)},
    \label{relError}
  \end{align} 
  is as small as desired.
  If the evaluation of \eqref{relError} is done with a low-precision
  approximation $\tilde{x}_m$ of $x_m$, the series might start to grow
  before the precision of $\epsilon_{\rm{rel}}$ is achieved, i.e.,
  $n'<N$.  In this case the calculation has to be repeated with a
  better approximation $\tilde{x}_m$, i.e., with a smaller error $\delta$.
  The expressions (\ref{norm1b}) and (\ref{scalar}) can be treated analogously.

  In Fig.~\ref{fig_Conv} we illustrate the asymptotic behavior of the
  sum in \eqref{norm1a}.  We want to evaluate it up to double
  precision (i.e., by demanding $\epsilon_{\text{rel}}$ $=$ $10^{-15}$) which is in
  general the required precision for possible further computations.  
  We evaluate \eqref{norm1a} for two different values
  of $\tilde{x}_m$.  The first value is calculated with double
  precision and
  $\delta=10^{-10}$.  Although a
  computation of the norm is possible for the lower part of the
  spectrum, for the twentieth eigenvalue $x_{20}$ the error $\epsilon_{\text{rel}}$
  does not become as small as the desired accuracy $10^{-15}$, 
  but starts to grow before.
  In a second step we evaluate the norm with an approximation
  $\tilde{x}_m$ obtained in a high (50 digit) precision calculation
  ($\delta=10^{-30}$) \cite{footnote}. This setup is depicted by the
  solid lines in Fig.~\ref{fig_Conv}(c) which cross the horizontal dashed line, 
  marking $\epsilon_{\text{rel}}=10^{-15}$,
  also for $x_{20}$. 
  However, the coefficients $K_n(x_m)$ in panel (b) are almost identical
  for the \mbox{high-} and low-precision calculations for $n<n'$, 
  i.e., before $\epsilon_{\rm{rel}}$ starts to grow. This
  observation serves as an \textit{a posteriori} confirmation of the
  asymptotic validity of \eqref{norm1a}. In the regime where the
  series is asymptotically meaningful
  $\abs{K_n} \lesssim \frac{1}{\sqrt{n!}}$ holds, while for $n>n'$,
  $\abs{K_n}$ decays exponentially,
  which is too slow to compensate the growth of $\sqrt{n!}$.
  These regimes are separated
  by clear kinks in Fig.~\ref{fig_Conv}(b). 

  The problems that arise when choosing a low-precision calculation
  are related to the structure of the three-term recursion \eqref{eqkn} that
  determines the sequence $K_n(\tilde{x}_m)$. For high values of
  $\tx_m$, the absolute value of $K_n(\tilde{x}_m)$ becomes so high
  that the sign-changing behavior of the sequence leads to large
  numerical errors.  For this reason, in a double-precision calculation even the
  spectrum cannot be determined for $x\gtrsim40$
  as the computation of 
  the function in Eq.~\eqref{gfunction} becomes numerically unstable. 
  The results of Sec.~\ref{sec_results} are all
  calculated using the high-precision setup of Fig.~\ref{fig_Conv}.

  \section{Absolutely convergent representation of $\p_{1,2}(x,z)$}
  \label{sec_convRep}

  The reason for the divergence of (\ref{norm1},\ref{scalar}) is that
  $\p_{1,2}(x,z)$ in \eqref{phi} is ill-defined for $|z+g|\geq 2g$ if
  $x\neq x_m$.  In the following, we present an analytical
  continuation of $\p_{1,2}(x,z)$ to the left complex half-plane
  $\Re(z)<0$ that is well defined for all $x$.  Using this analytical
  continuation, we perform the integration of the scalar product
  \eqref{scalarProd} over the full complex plane and obtain finite
  values for the norm \eqref{norm1} and scalar products \eqref{scalar}
  also for $x\neq x_m$.  For $x = x_m$ this evaluation yields the
  values associated to the true eigenstates.  A comparison with the
  results of the preceding subsection then establishes the asymptotic
  validity of formulae (\ref{norm1},\ref{scalar}).

  The squared norm $\langle\psi_m|\psi_m\rangle$ reads
  \begin{align}
    \langle\psi_m|\psi_m\rangle=
    \int\frac{\rd z\rd\bz}{\pi} e^{-z\bz}\psi(x_m,\bz)\psi(x_m,z).
    \label{norm2}
  \end{align}
  To compute (\ref{norm2}), one needs $\psi(x_m,z)$ in the whole
  complex plane. Using the analytical continuation of $\p_1(z),
  \p_2(z)$ into the left complex half-plane, we define for arbitrary
  real $x$,
  \begin{multline}
    \langle\psi(x)|\psi(x)\rangle 
    = \int\limits_{\text{left\ HP}}\frac{\rd z\rd\bz}{\pi}
    e^{-z\bz}\p_1(x,\bz)\p_1(x,z)
    \\
    +
    \int\limits_{\text{right\ HP}}\frac{\rd z\rd\bz}{\pi} 
    e^{-z\bz}\p_2(x,-\bz)\p_2(x,-z)
    \\
    =  \int\limits_{\text{left\ HP}}\frac{\rd z\rd\bz}{\pi} 
    e^{-z\bz}\left(\abs{\p_1(x,z)}^2 + \abs{\p_2(x,z)}^2 \right).
    \label{norm3}
  \end{multline}
  The function $\psi(x,z)$, defined by $\p_1(x,z)$ in the left and by
  $\p_2(x,-z)$ in the right half-plane, exhibits a discontinuity along
  the imaginary axis if $x-g^2\notin$ spec$H_+$, but becomes analytic
  everywhere if $x=x_m$.  By replacing (\ref{norm2}) by (\ref{norm3}),
  we obtain finite expressions for all $x$ and the correct ones if
  $x-g^2$ belongs to the spectrum.

  \begin{figure}[t!]
    \centering
    \includegraphics[width=0.9\hsize,clip]{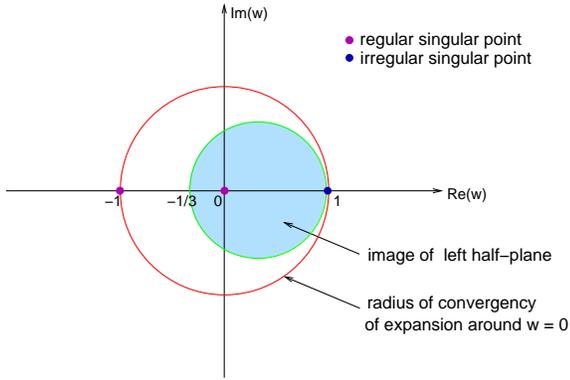}
    \vspace*{-0.3cm}
    \caption{Image of left $z$ half-plane (light blue) in the $w$ plane
      using the M\"obius transformation \eqref{moebius}.}
    \label{w-circle}
  \end{figure}

  The analytic continuation of $\p_{1,2}(x,z)$ from the circle with
  radius $2g$ around $z=-g$ to the left half-plane is constructed with
  a M\"obius transformation. The system of ordinary differential
  equations satisfied by $\p_1(z),\p_2(z)$ reads \cite{db}
  \begin{subequations}%
    \begin{align}%
      (z+g)&\p_1'(z) +(gz-x+g^2)\p_1(z) +\D\p_2(z) = 0,\\
      (z-g)&\p_2'(z) -(gz+x-g^2)\p_2(z) +\D\p_1(z) = 0.
    \end{align}%
    \label{coup-sys}%
  \end{subequations}%
  Primes denote differentiation with respect to  $z$.
  The system is characterized by two regular singular points at $z=\pm g$ and
  one irregular singular point at infinity~\cite{ince}. Consider now the
  transformation, $z \rightarrow w(z)$,
  \begin{align}
    w(z)=\frac{z/g+1}{z/g-3},\quad z(w)=g\frac{3w+1}{w-1},
    \label{moebius}
  \end{align}
  which maps the irregular singular point at $z=\infty$ to $w=1$ and
  the two regular singular points at $z=\mp g$ to $w=0$ and $w=-1$
  respectively (see Fig.~\ref{w-circle}).  The left $z$ half-plane is
  mapped onto the disk with radius $2/3$ and center $w=1/3$ whereas
  the imaginary $z$-axis maps onto the boundary of the disk.  With
  $\tx=4g^2-x$, the system (\ref{coup-sys}) reads in the variable $w$,
  \begin{subequations}%
    \begin{align}%
      &w(w-1)\tp_1'(w)
      -
      \big(\tfrac{4g^2}{w-1}+\tx\big)\tp_1(w)
      =
      \D\tp_2(w),
      \\
      &\tfrac{w^2-1}{2}\tp_2'(w) 
      +
      \big(\tfrac{4g^2}{w-1}+6g^2-\tx\big)\tp_2(w)
      =
      \D\tp_1(w).
    \end{align}%
    \label{coup-sys-w}%
  \end{subequations}%
  It has regular singular points at $w=-1,0$ and one irregular
  singular point at $w=1$, while $w=\infty$ is a regular point.  Now
  we may expand the functions $\tp_{j}(w)$ with $j=1,2$ in a power
  series around the regular singular point $w=0$,
  \begin{align}
    \tp_j(x,w)=\sum_{n=0}^\infty a_n^{(j)}(x)w^n, 
    \label{pot}
  \end{align}
  and obtain, with the definition
  \begin{align}
    \v_{n+1} =\left(
      \begin{array}{l}
        a^{(2)}_{n+1}\\
        a^{(1)}_{n\phantom{+1}}
      \end{array}
    \right),
  \end{align} 
  for $n\ge-1$ the following four-term matrix recurrence
  \begin{align}
    \v_{n+1}=\A_n\v_{n}+\B_n\v_{n-1}+\C_n\v_{n-2}.
    \label{mrecur}
  \end{align}
  The coefficient matrices are given by
    \begin{align}%
      \A_n&=
      \left(
        \begin{array}{cc}
          \frac{(4g^2-2\tx+n)(n-x)+2\D^2}{(n+1)(n-x)}
          &
          \frac{2\D}{n+1}(1-\frac{\tx+2n-2}{n-x})
          \\[1ex]
          \frac{-\D}{n-x}
          &
          \frac{\tx+2n-2}{n-x}
        \end{array}
      \right)\!,\nonumber\\
      \B_n&=
      \left(
        \begin{array}{cc}
          \frac{(2\tx-12g^2+n-1)(n-x)-2\D^2}{(n+1)(n-x)}
          &
          \frac{2\D(n-2)}{(n+1)(n-x)}
          \\[1ex]
          \frac{\D}{n-x}
          &
          \frac{2-n}{n-x}
        \end{array}
      \right)\!,\nonumber\\
      \C_n&=
      \left(
        \begin{array}{ll}
          \frac{2-n}{n+1} & 0
          \\[1ex]
          0 & 0
        \end{array}
      \right)\!.%
    \end{align}%
  The condition $\p_j(x,z)=\tp_j(x,w(z))$
  fixes the initial values of $\v_0$ and $\v_1$, namely
  \begin{subequations}%
    \begin{align}%
      a_0^{(2)}&=e^{g^2},~~
      a_0^{(1)}=\frac{\D}{x}e^{g^2},
      \\
      a_1^{(2)}&=2\Big(x-2g^2-\frac{\D^2}{x}\Big)e^{g^2}.
    \end{align}%
  \end{subequations}%
  Together with $\v_{-1}=\mathbf{0}$, all $\v_n$ for $n\ge 2$ follow
  recursively.  

  The asymptotic behavior of the $a_n^{(j)}$ can be inferred from
  (\ref{mrecur}) as $\lim_{n\ra\infty}$
  $|{a_{n+1}^{(j)}}/{a_n^{(j)}}|$ $=$ $1$. Hence the radius of
  absolute convergence of the expansion (\ref{pot}) is 1 for all $x$,
  as expected from the singularity structure of (\ref{coup-sys-w}):
  The expansion around $w=0$ converges up to the neighboring singular
  points, $w=\pm1$.  Now we can perform the integral in (\ref{norm3})
  because the image of the left half-plane is contained entirely
  within the region of absolute convergence of the series expansion
  (\ref{pot}), which furnishes the desired analytical continuation of
  the functions $\p_j(x,z)$. The integral \eqref{norm3} reads in $w$ coordinates 
  \begin{multline}
    \langle\psi(x)|\psi(x)\rangle
    =
    \int\limits_{\text{disk}} \frac{\rd w \rd\bar w}{\pi}
    \frac{4 g^2}{\abs{w-1}^4}
    e^{-z(w)z(\bar w)}
    \\\times
    \left(\abs{\tilde\p_1(x,w)}^2 + \abs{\tilde\p_2(x,w)}^2 \right).
    \label{normWrep}
  \end{multline}

\begin{figure}[t!]
    \centering
    \includegraphics[width=0.8\hsize,clip]{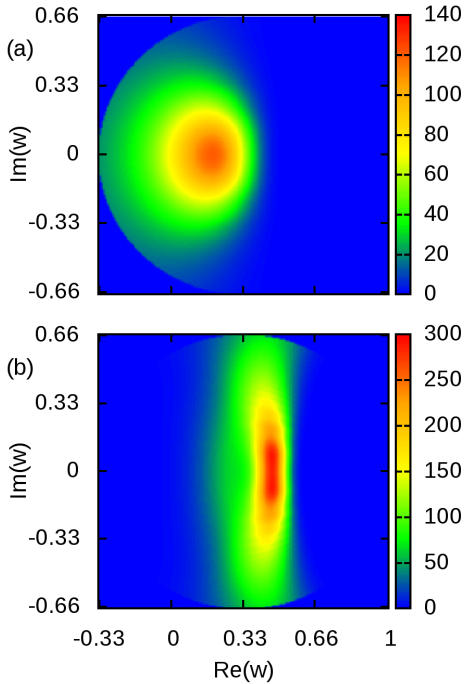}
    \vspace*{-0.3cm}
    \caption{The integrand of \eqref{normWrep} for the first (a) and
      the fifth (b) eigenenergy ($x_0=0.06038...$ and $x_5=4.9355...$) in
      the positive parity subspace for parameters $g/\omega=0.7$ and
      $\D/\omega=0.25$.  \label{fig_Integrand}}
  \end{figure}

  The numerical integration in \eqref{normWrep} is done in polar
  coordinates over the disk in the $w$ plane obtained by mapping
  the left half-plane with the M\"obius transformation \eqref{moebius} (see Fig.~\ref{w-circle}). 
  Fig.~\ref{fig_Integrand} shows that the integrand in \eqref{normWrep} is
  a smooth function, for which the numerical integration is easy to control.
  The singularity that is still present in the wave function $\tp_j(x,w)$ for $w=1$ 
  is suppressed in the integrand by the regularizing factor $e^{-z(w)z(\bar w)}$. As expected 
  from the convergence properties of $\tp_j(x,w)$, the representation of the norm in Eq.~\eqref{normWrep}
  depends smoothly on the argument $x$. This means that for values of  $x$ that differ only slightly from the
  spectral values used in Fig.~\ref{fig_Integrand}, the corresponding plots of the integrand are essentially the same.

  Finally, we confirmed the validity of the asymptotic formula \eqref{norm1approx} 
  by comparing it to a numerical integration of \eqref{norm3}. 
  The validity of the remaining expressions (\ref{norm1},\ref{scalar}) was also
  checked.

  \section{Spin observables}
  \label{sec_results}

  In this section we calculate the time evolution of
  $\expec{\sigma_z(t)}$ and $\expec{\sigma_x(t)}$ using the method
  presented in Sec.~\ref{sec_timeEvol}.  The ``population inversion'' 
  $\sigma_z(t)$ has been thoroughly investigated for the resonant case 
  ($g/\omega\ll 1$ and $\D/\omega\simeq 0.5$) within the
  rotating-wave approximation, where it shows the typical collapse and
  revival oscillations~\cite{eberly} as well as within
  analytical~\cite{feranchuk} and numerical~\cite{larson} treatments
  of the full model.  
  Here we follow a different path by investigating the problem 
  for the non-resonant case and arbitrary parameters in all coupling regimes.
  We find a qualitatively different behavior for $\expec{\sigma_z(t)}$ and $\expec{\sigma_x(t)}$ 
  for strong coupling.
  As will be discussed below, this difference can be understood in terms of the
  $\Zz_2$ symmetry, manifested in the two invariant parity chains.

  \subsection{Time evolution of $\expec{\s_z}$}

\begin{figure}[t!]
    \centering
    \includegraphics[width=\hsizeNew]{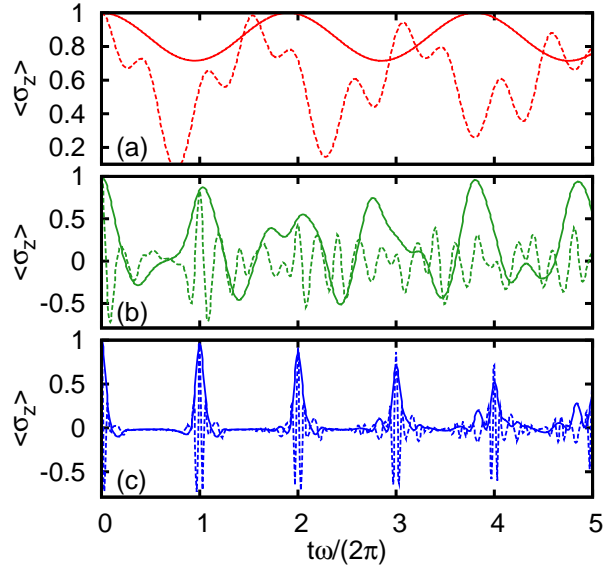}
    \vspace*{-0.3cm}
    \caption{$\expec{\s_z(t)}$ for $\D/\omega=0.25$ and (a) $g/\omega=0.1$, (b)
      $g/\omega=0.7$, (c) $g/\omega=2$.  The evolution starts with coherent states
      with $\alpha=0$ (solid line) and $\alpha=2$ (dashed line).}
    \label{fig_sigmaZ}
  \end{figure}

  To study the time evolution of $\s_z$, we prepare the system at
  time $t$ $=$ $0$ in a product of a coherent state $\ket{\a}= e^{-\a^2/2+\a z}$ with 
  an eigenstate of $\s_z$
\begin{align}
\ket{\psi(0)} = \ket{\a} \otimes \ket{+1}.
\label{iniz}
\end{align} 
  In terms of parity chains it reads, 
  \begin{align}
    \ket{\psi(0)} = e^{-\a^2/2} \big(|\ch(\a z),+\rangle+|\sh(\a z),-\rangle\big).
  \end{align}
  Using the isomorphism \eqref{isoP}, we find for the time evolution of $\expec{\s_z}$
\begin{multline}
\expec{\s_z(t)} =
e^{-\a^2}[\bra{\cosh(\a z)}e^{-iH_+t} \T e^{iH_+t} \ket{\cosh(\a z)}
\\
- \bra{\sinh(\a z)}e^{-iH_-t} \T e^{iH_-t} \ket{\sinh(\a z)}] ,
\label{sigmaz}
\end{multline} 
  where $\T$ is the reflection operator in $\cal B$.  This expectation
  value is plotted in Fig.~\ref{fig_sigmaZ}.
  As can be seen from \eqref{sigmaz} the time evolution is solely governed by the 
  commutator $[\T,H_{\pm}]$ and the system evolves separately in each parity
  subspace. Thus, when evaluating
  the expression, one can assign a definite parity to the frequencies $E^\pm_n-E^\pm_m$.
  As these frequencies are always of order $(n-m)\om$ for
  sufficiently strong coupling $g/\D$, the behavior on time scales of a few
  $2\pi/\om$ is dominated by $\om$ as the smallest frequency.

  \subsection{Time evolution of $\expec{\s_x}$}

\begin{figure}[t!]
    \centering
    \includegraphics[width=\hsizeNew]{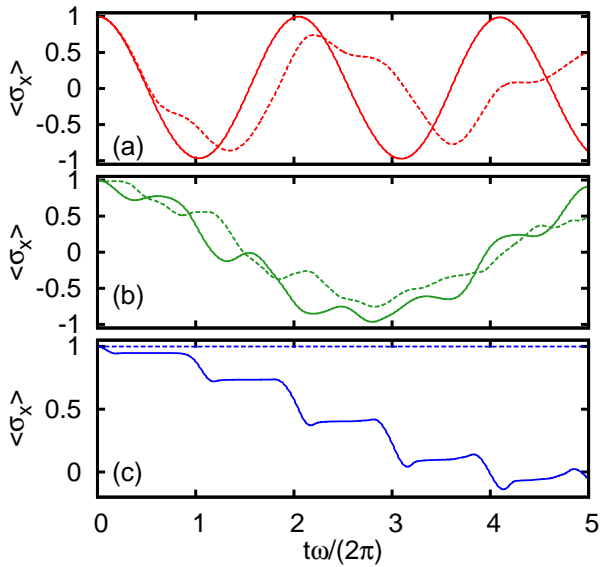}
    \vspace*{-0.3cm}
    \caption{$\expec{\s_x(t)}$ for the same setup as in
      Fig.~\ref{fig_sigmaZ}.  Note that (c) shows that
      $\expec{\s_x(t)} \simeq 1$ for all times if the initial state is
      given by the linear combination of two cat states in
      Eq.~\eqref{lcCat}. We observe a sharp step-like decay (solid line) in (c) for
      a system prepared in the vacuum ($\alpha=0$). For a system
      prepared in a coherent state with $\alpha=g/\omega=2$, a constant plateau is seen in (c) (dashed line).
      \label{fig_sigmaX}}
  \end{figure}

  In complete analogy to the case of $\s_z$, we prepare the system at
  time $t$ $=$ $0$ in a product of a coherent state with an eigenstate of $\s_x$
\begin{align}
\ket{\psi(0)} = \ket{\a}\otimes \frac{1}{\sqrt{2}} (\ket{+1}+\ket{-1}).
\label{inix}
\end{align} 
  Using the representation in the two parity subspaces, we obtain the
  time evolution of $\expec{\s_x}$ as
\begin{align}
\expec{\s_x(t)} =\frac{1}{2}
\bra{\a}e^{-iH_+t} e^{iH_-t}
+ e^{-iH_-t} e^{iH_+t} \ket{\a}.
\label{sigmax}
\end{align} 
  This expectation value is plotted in Fig.~\ref{fig_sigmaX}.
  When evaluating the expression in \eqref{sigmax}, 
  the frequencies have the form $E_n^--E_m^+$ with eigenenergies from {\it
  different} parity chains. These frequencies may become arbitrary small. In
  particular the frequency $E_n^+-E_n^-$ is small for
  strong coupling, because $E_n^+$ and $E_n^-$ both approach the same value $n\om-g^2/\om$ 
  for growing $g/\D$ from above and below. Therefore small frequencies
  dominate $\langle\s_x(t)\rangle$ for sufficiently strong coupling.
  We note that the frequency $E_n^+-E_n^-$ vanishes if the parameters $g$ and $\Delta$
  satisfy the Juddian condition $K_n(n\om)=0$.

  For $g/\omega=2$ and  $\D/\omega=0.25$ 
(i.e., in the so-called deep strong coupling regime \cite{sol})
we observe  
two features in Fig.~\ref{fig_sigmaX}(c). The first is the sharp step-like
  decay of $\expec{\s_x(t)}$ for a system prepared in the vacuum ($\a=0$) at $t=0$. Each  
   step  has the length $2\pi/\om$. This
  structure disappears for higher values of $\a$.

  Secondly,  $\expec{\s_x(t)}$ stays perfectly constant 
  for $\a=g/\om$ as depicted by the dashed line in Fig.~\ref{fig_sigmaX}(c).
  In this case, the representation of the initial state \eqref{inix} in the parity subspaces
  is a superposition of the two ``Schr\"odinger cat states'' $\ket{C_\pm}\in{\cal
    H}_\pm$ \cite{cat-ref},
  \begin{align} \label{lcCat}
    |\Psi(0)\rangle&=
    \frac{1}{\sqrt{2}}(\ket{C_+}+\ket{C_-}) 
    \text{~~for~~}\a=\frac{g}{\om},    
    \\
    \ket{C_+}=&
    e^{-g^2/2\om^2}\big(\ch(gz/\om)\otimes|+1\rangle +\sh(gz/\om)\otimes|-1\rangle\big) 
    ,\nonumber
    \\
    \ket{C_-}=&
    e^{-g^2/2\om^2}\big(\ch(gz/\om)\otimes|-1\rangle+\sh(gz/\om)\otimes|+1\rangle\big)
    .\nonumber
  \end{align}

  The states $\ket{C_\pm}$ are eigenstates of $H_\pm$ for $\D=0$ and one would expect
  their approximate conservation over time for small $\D$ but decay at longer time scales
  or  appreciable ratio $\D/g$. It is therefore surprising that the dynamical protection of
  $\ket{C_\pm}$ (and concomitantly $\s_x$) extends to long times (see next section) and applies
  to quite large $\D/g$, which is 0.125 in Figs.~\ref{fig_sigmaX}(c) and \ref{fig_sigmaxL}(c).
  This remarkable feature of the QRM at couplings $g/\om >1$ is related to the excellent
  approximation of the true ground state in each parity chain by the shifted vacuum state
  (see Eq.~\eqref{shift-osc})
  and will be discussed in more detail elsewhere. The dynamical protection of $\ket{\Psi(0)}$
  leads to the constant plateau (dashed line) in Fig.~\ref{fig_sigmaX}(c). Fig.~\ref{fig_sigmaxL}(c)
  shows that this behavior persists even on long time scales.
  
  \subsection{Long-time behavior}

  In the preceding subsections we pointed out that for sufficiently high $g/\D$
  the frequencies appearing in the time evolution of $\expec{\s_z(t)}$ have an
  approximate lower bound of $\om$ while this is not the case for $\expec{\s_x(t)}$. This is inferred
  from an analysis of the representation in the parity subspaces and can be seen
  by comparing Figs.~\ref{fig_sigmaZ}(b),(c) with Figs.~\ref{fig_sigmaX}(b),(c). Only in the latter
  slow oscillations are visible.

  \begin{figure}[t!]
    \centering
    \includegraphics[width=\hsizeNew]{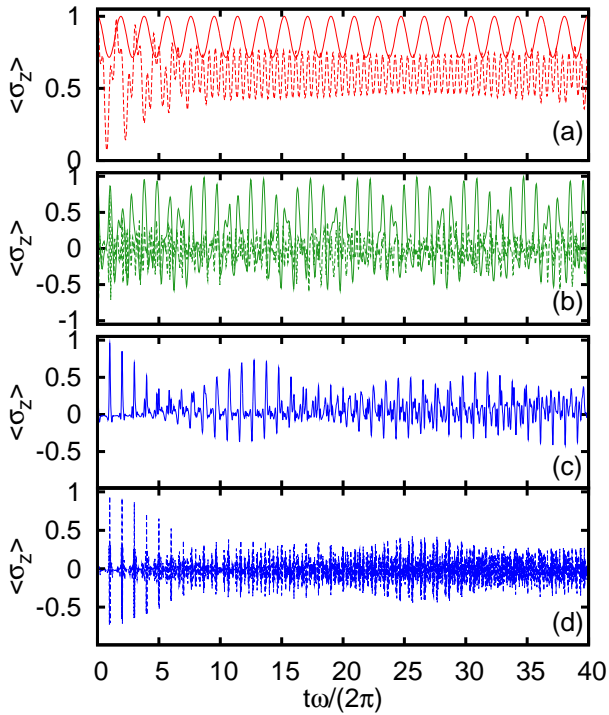}
    \vspace{-0.3cm}
    \caption{ $\expec{\s_z(t)}$ for times up to 40 fundamental periods
      ($T_0=2\pi/\om$) and the same parameters as in
      Fig.~\ref{fig_sigmaZ}. 
      The case $g/\om=2$ is split into panels (c), $\a=0$, and (d), $\a=2$, 
      for clarification.
      The effect of the modulating beat
      frequencies is not visible for $g/\omega=0.1$ on this time scale, 
      whereas it appears clearly for $g/\om=0.7$ and $2$. In all three cases 
      higher harmonics dominate the time evolution for $\a=2$
      (dashed lines in (a) and (b), resp. panel(d)).}
    \label{fig_sigmazL}
  \end{figure}
  \begin{figure}[t!]
    \centering
    \includegraphics[width=\hsizeNew]{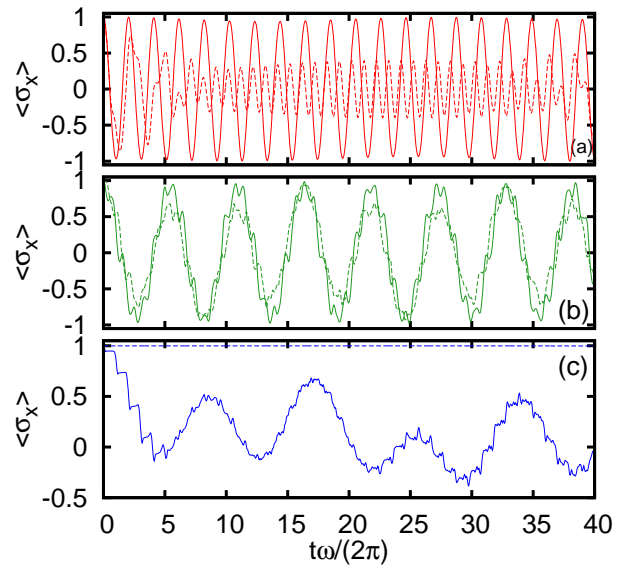}
    \vspace{-0.3cm}
    \caption{$\expec{\s_x(t)}$ for times between 0 and $40T_0$ as in
      Fig.~\ref{fig_sigmazL} and parameters as in
      Fig.~\ref{fig_sigmaX}. The qualitative behavior depends clearly
      on the initial state for small coupling $g/\omega=0.1$ (top panel) but
      not for $g/\omega=0.7$ (b).  The dominating period $T^*>T_0$
      grows with $g$ due to decreasing differences $E_n^+-E_n^-$. For
      $\a=g/\omega=2$, $\expec{\s_x}$ stays constant.\label{fig_sigmaxL}}
  \end{figure}

  This effect is also observed in the long-time behavior
  displayed in Figs.~\ref{fig_sigmazL} and~\ref{fig_sigmaxL}.
  For weak coupling ($g/\omega=0.1$) the time evolution of 
  $\expec{\s_z(t)}$ and $\expec{\s_x(t)}$ is rather similar in Figs.~\ref{fig_sigmazL}(a) and~\ref{fig_sigmaxL}(a).
  For strong coupling ($g/\omega \geq 0.7$) the same analysis as for short time scales applies.
  In the case of $\s_z$, $\om$ remains the smallest frequency and the
  oscillations are dominated by higher harmonics, although a modulation
  by small beat frequencies, due to the superposition of contributions 
  from both parity subspaces \eqref{sigmaz},  
  is visible.
  By contrast, the overall behavior of $\expec{\s_x(t)}$ for $g/\omega \geq 0.7$ is dominated
  by oscillations with period $T^*>2\pi/\om$, as seen in
  Fig.~\ref{fig_sigmaxL}(b) and (c).  According to the argument of the
  previous subsection, $T^*$ grows with $g$.
  For $g/\omega=0.7$
  the dominant frequency
  $1/T^*$ is then essentially the same for initial coherent states with
  $\a=0$ and $\a=2$, see Fig.~\ref{fig_sigmaxL}(b). This is true
  also for the case of $g/\omega=2$, where it is seen that the step-like structures
  follow an overall oscillation with period $T^*$. 
  Only in the special case of $\a=g/\omega$, in which the time evolution is dynamically protected,
  the initial expectation value is maintained for
  arbitrary long times (dashed line in Fig.~\ref{fig_sigmaxL}(c)).

  \section{Conclusion}

  We have described a method to compute exactly all eigenstates of the
  quantum Rabi model together with their norms and scalar products
  with states of the standard basis. In this way the development of an
  arbitrarily prepared initial state can be obtained for short and
  long times. The method uses the $\Zz_2$ invariance of the model, the
  representation of the oscillator degree of freedom in the Bargmann
  space of analytical functions and a conformal mapping to obtain
  absolutely convergent expansions of the norms and overlaps with Fock
  states. 
  The method overcomes the limitations of numerical procedures or 
  analytical approximations based on a truncated Hilbert space.
  Moreover, these techniques themselves
  can now be mathematically justified by comparison with the analytical solution
  and retain in this way their usefulness for practical calculations.
  This will be discussed elsewhere. 
  The splitting of the full space into two invariant subspaces
  results in a qualitatively different behavior of $\expec{\s_z(t)}$ and
  $\expec{\s_x(t)}$ on intermediate and long time scales: 
  $\expec{\s_z(t)}$ is always dominated by integer multiples of the
  photon frequency $\om$. These oscillations are modulated by small beat frequencies
  because the QRM spectrum is not equidistant and both parity chains interfere.
  $\expec{\s_x(t)}$, on the other hand, shows oscillations with a dominant frequency 
  much smaller than $\om$, which is directly related to the mixing of both parity chains
  effected by the operator $\s_x$. The oscillation period grows with growing coupling
  between qubit and radiation field. For initial coherent states fulfilling the condition
  \eqref{lcCat}, we observe  ``dynamical protection'' in the deep strong coupling regime 
  which could be relevant to the control of fast quantum gates within
  circuit QED.

  \begin{acknowledgments}
    This work was supported in part by Deutsche Forschungsgemeinschaft through TRR~80.
  \end{acknowledgments}

\end{document}